\documentclass[final]{IEEEtran}
\usepackage{latexsym,pictex,graphics}
\newtheorem{theorem}{Theorem}

\newtheorem{proposition}{Proposition}
\newtheorem{lemma}{Lemma}

\newcommand{\fat}[1]{\mbox{\boldmath$#1$}}
\newcommand{\sfat}[1]{\mbox{\boldmath$\scriptstyle#1$}}
\newcommand{\nfat}[1]{\mbox{\rlap{$#1$}\kern.2pt\rlap{$#1$}\kern.2pt$#1$}}

\def\endtheorem{\hspace*{\fill}~$\Box$\par\endtrivlist\unskip}

\title{A Formulation of the Channel Capacity of Multiple-Access
Channel}

\author{\authorblockN{Yoichiro Watanabe\thanks{Yoichiro Watanabe is with the
Department of Intelligent Information Eng. and Sci., Doshisha University,
Kyotanabe, Kyoto, 619-0321 Japan.}, \IEEEmembership{Member, IEEE} and
\authorblockN{Koichi Kamoi\thanks{Koichi Kamoi is with Knowledge X Inc.,
Wako-Shi, Saitama, 351-0114 Japan.}, \IEEEmembership{Member, IEEE}}}}

\begin{document}

\maketitle

\begin{abstract}
  The necessary and sufficient condition of the channel capacity is
  rigorously formulated for the $N$-user discrete memoryless
  multiple-access channel (MAC). The essence of the formulation is to
  invoke an {\em elementary} MAC where sizes of input alphabets are not
  greater than the size of output alphabet. The main objective is to
  demonstrate that the channel capacity of an MAC is achieved by
  an elementary MAC included in the original MAC. The proof is quite
  straightforward by the very definition of the elementary MAC. Moreover
  it is proved that the Kuhn-Tucker conditions of the elementary MAC are
  strictly sufficient and obviously necessary for the channel capacity.
  The latter proof requires some steps such that for the elementary MAC
  every solution of the Kuhn-Tucker conditions reveals itself as local
  maximum on the domain of all possible input probability distributions
  and then it achieves the channel capacity. As a result, in respect of
  the channel capacity, the MAC in general can be regarded as an aggregate
  of a finite number of elementary MAC's.
\end{abstract}


\section{Introduction}
\PARstart{T}{he} channel capacity is without question recognized as an
essential subject of the (discrete memoryless) multiple-access channel
(MAC) with $N$ input-terminals and one output-terminal. Since it is
defined as the maximum of the mutual information, we are familiar with
the so-called Kuhn-Tucker conditions as necessary to achieve the channel
capacity. Up to now, however, the Kuhn-Tucker conditions are not
entirely examined as sufficient for the $N$-user MAC except for the
simplest case of single user discrete memoryless channel (DMC). Thus it
is natural to ask how the sufficiency could be formulated for the case
of MAC in general.

In this paper, we demonstrate that there exists a non-trivial MAC where
the Kuhn-Tucker conditions are strictly sufficient (and obviously
necessary) for the channel capacity. We refer to it as {\em an elementary}
MAC whose sizes of input alphabets are not greater than the size of output
alphabet. Evidently the DMC is an elementary MAC. The most of this paper
is devoted to the proof that the Kuhn-Tucker conditions are sufficient
(the necessity is self-evident) for the channel capacity of the elementary
MAC.

On the other hand, for any given $N$-user MAC we can uniquely determine a
finite set of elementary MAC's. It is an aggregate of \emph{the largest}
possible elementary MAC's included in the given $N$-user MAC and is
referred to as {\em the master elementary set} to be denoted by
$\Omega_N$. We demonstrate that the channel capacity of the $N$-user MAC
is achieved by the channel capacity of \emph{an elementary} MAC of the set
$\Omega_N$. The proof here appears quite straightforward by merely
appealing to the very definition of the elementary MAC without asking for
any other features such that the Kuhn-Tucker conditions are sufficient.

Thus an MAC in general can be regarded as simply an aggregate of
elementary MAC's where the Kuhn-Tucker conditions are necessary and
sufficient for the channel capacity. Roughly speaking, an MAC comprises a
finite number of elementary MAC's. This statement is a basic idea behind
our formulation of this paper.

Here we must emphasize that several steps are required to prove the
sufficiency of the Kuhn-Tucker conditions of the elementary MAC. In fact,
we need to prove two distinctive features: The first is that for the
elementary MAC every solution of the Kuhn-Tucker conditions is {\em local
maximum} on the domain of all possible input probability distribution
(IPD) (or the probability simplex, see Cover~\cite{ELEMENTS} for this
terminology and we refer to them as IPD {\em vectors} for our purposes).
The second is that for the elementary MAC a set of IPD vectors for which
the value of the mutual information is not smaller than the arbitrary
positive number is {\em connected} on the domain of all possible IPD
vectors. To prove the second property of connectedness we require the
first property of local maximum. Then it follows after a bit of procedures
that solutions of the Kuhn-Tucker conditions are uniquely determined, that
is, each solution takes the same value for the mutual information and
therefore it achieves the channel capacity.

For the explicit description of our concept we take a logical stream as
follows: After defining the elementary MAC and determining the set
$\Omega_N$, we first prove as the main theorem that the channel capacity
of an $N$-user MAC is achieved by the channel capacity of an elementary
MAC of $\Omega_N$ and then we prove as the second theorem that the
Kuhn-Tucker conditions are sufficient for the channel capacity of the
elementary MAC. These are the main objective of this paper.

After Shannon~\cite{Two-Way}, the study of multiuser channel
(multiterminal network) has long been carried out in various fields
including MAC, broadcast channel, relay channel, interference channel,
two-way channel and so forth. The channel coding theorem was proved
independently by Liao~\cite{Liao}, Ahlswede~\cite{Ahlswede} and
Meulen~\cite{Meulen}. These are followed by many authors
(\cite{Ulrey,Gaarder,CHANG,Cover,Hui,VERDU,KRAMER}) to provide a deeper
insight into the capacity region. Recently, information-theoretic approach
has been adopted to large scale networks, such that code division
multiple-access channel, continuous time multiple-access channel and
space-time multiple-access channel (e.g.,
\cite{SHAMAI-I,SHAMAI-II,SUARD,WEI-YU,GUPTA}), as we know. Also a
computation procedure for the channel capacity of MAC has been developed
(e.g., \cite{Rezaeian-Grant}).

The purpose of the study of MAC is mostly to investigate the
multiuser coding that retains both reliability and efficiency. The
investigation has been carried out mostly on the computational
calculations for practical applications. Not much has been made for
the mathematical rigorousness of the formulation since it appears
rather hard to solve a non-linear optimization problem of the mutual
information with several variables under constraints. We have been
highly expecting a theoretical foundation, in particular, for the
rigorous evaluation of the channel capacity and the exact
determination of the capacity region for the MAC in general. These
can provide us with the mathematical essence as well as the fine
structure inherent in the MAC. Also we believe that these can in
part complement the computational approaches to various applications
as well.

In the past, for the MAC of two-user and binary output~\cite{IT-42-5}, we
have shown that the Kuhn-Tucker conditions are necessary and sufficient
for the channel capacity. The basic idea was to identify the channel
matrix of the MAC as {\em a linear} mapping from the convex-closure of IPD
vectors to the range of output probability distributions. Now we expand
the idea and remind a clear conception to describe the MAC as a pair of
channel matrix $P$ and domain $X$ where $X$ is a set of IPD vectors and
$P$ is interpreted as a mapping ({\em non-linear} in general) from IPD
vectors to output probability distributions. Any quantity such as mutual
information and so forth is considered as a function of IPD vectors
defined on a restricted domain (a sub-set) of $X$. These are seemingly
non-standard in contrast to the ordinary description of information theory
as in~\cite{ELEMENTS}. However, we assure ourselves that these conceptions
including the notation adopted in this paper are so successful to overcome
some difficulties and cumbersome procedures underlying in the non-linear
optimization problem relating to the mutual information of the MAC.

In Section~\ref{S:E-MAC} we describe some expressions and definitions to
be used in this paper. In particular we introduce an elementary MAC and
the master elementary set for the MAC. In Section~\ref{S:Main} we prove
the main theorem of this paper as Theorem~\ref{T:C(F)} followed by
indicating the value of this theorem. In Section~\ref{S:BEQ} we
investigate distinctive features of the elementary MAC that are required
to prove the succeeding Theorem~\ref{T:E-MAC}. In
Section~\ref{S:BinaryMAC} we investigate an special case of binary-inputs
MAC. In Section~\ref{S:PROOF} we prove Theorem~\ref{T:E-MAC}. In the last
Section~\ref{S:CONCLUTIONS} we summarize the paper with some comments.

\section{Elementary MAC} \label{S:E-MAC}
In this section we introduce an elementary MAC with some expressions and
definitions to be used in this paper.

{\it An $N$-user MAC} is specified by $N$ input alphabets $A_k$ with
size of $n_k$, $k=1,\cdots,N$, an output alphabet $B$ with size of $m$,
and an $m$ by $(n_1\times\cdots\times n_N)$ channel matrix
$P=[P(j|i_1,\cdots,i_N)]$ of transmission probabilities
$P(j|i_1,\cdots,i_N)$'s to be given {\it a priori} for the MAC, where
$j$=$0,\cdots,m-1$, $i_k$=$0,\cdots,n_k-1$, and $\sum_{j=0}^{m-1}$
$P(j|i_1,\cdots,i_N)=1$. Assume that there is no zero row in $P$,
$n_k\geq 2$, $m\geq 2$, and transmission is synchronized. The model thus
defined is called an $N$-user MAC with a type $(n_1,\cdots,n_N;m)$.

{\it The IPD vector} $\fat{p}_k$ is assigned to an $n_k$-tuple column set
of input probability $(p_k(0),\cdots,p_k(n_k-1))^{\tt T}$, $k=1,\cdots,N$,
where $(\cdots)^{\tt T}$ implies a transposition, defined on $A_k$ with
the probability constraint $\sum_{i_k=0}^{n_k-1} p_k(i_k)=1$. Thus each
$\fat{p}_k$ is located on an $(n_k-1)$-dimensional {\it simplex} $X_k$
with $n_k$ vertices $\fat{e}_{k\ell}$, $\ell=0,\cdots,n_k-1$. Here
$\fat{e}_{k\ell}$ is a unit column vector and takes $1$ in the $\ell$th
column component and $0$ elsewhere. Obviously each $X_k$ is convex and is
observed as \emph{a domain} of $\fat{p}_k$.

{\it The face} $F_k$ of $X_k$ is defined by an $(f_k-1)$-dimensional
simplex whose $f_k$ vertices are chosen from vertices
$\fat{e}_{k0},\cdots,\fat{e}_{k(n_k-1)}$ of $X_k$, where $f_k\leq n_k$.
A set of $f_k$ indices of $F_k$ is denoted by
$\Lambda({F_k})=\{\ell|\fat{e}_{k\ell}\in F_k\}$. There are several
choices for $f_k$ indices. Obviously $\Lambda(F_k)\subset\Lambda(X_k)$ and
$\Lambda(X_k)=\{0,\cdots,n_k-1\}$. Zero-dimensional faces are vertices,
one-dimensional faces are lines, and so forth. If an IPD vector
$\fat{p}_k$ is on the boundary of $X_k$, then there exists
\emph{a minimum} face $F_k$ which contains $\fat{p}_k$ exactly inside (and
not on the boundary of) $F_k$. Thus if $\fat{p}_k$ is $p_k(i_k)=0$ for
$i_k\not\in\Lambda({F_k})$ and $p_k(i_k)>0$ for $i_k\in\Lambda({F_k})$,
then $F_k$ for $\Lambda({F_k})$ is the minimum face which contains
$\fat{p}_k$ and is uniquely determined. If $p_k(i_k)>0$ for all $i_k$'s,
then the minimum face which contains the $\fat{p}_k$ is $X_k$ itself. Here
$F_k$ is also referred to as \emph{a sub-domain} of $X_k$. Also
$\fat{p}_k$, with $p_k(i_k)>0$ for $i_k\in\Lambda({F_k})$ and $p_k(i_k)=0$
for $i_k\not\in\Lambda({F_k})$, is naturally regarded as an
$(f_k-1)$-dimensional vector on $F_k$ even it is still an
$(n_k-1)$-dimensional vector on the whole domain $X_k$.

{\it The Kronecker product} of $\fat{p}_1$ and $\fat{p}_2$ is
defined here by
\[
\fat{p}_1\times\fat{p}_2\equiv \left[
\begin{array}{c}
p_1(0)\fat{p}_2 \\ \vdots \\ p_1(n_1-1)\fat{p}_2
\end{array}\right]
\]
and then the Kronecker product of $\fat{p}_1,\cdots,\fat{p}_k$ is defined
by induction: $\fat{p}_1\times\cdots\times\fat{p}_k$ $\equiv$
$(\fat{p}_1\times\cdots\times\fat{p}_{k-1})\times\fat{p}_k$,
$k=3,\cdots,N$. In the same way we arrange the Kronecker product of
$X_1,\cdots,X_N$ as
\[
X\equiv X_1\times \cdots \times X_N.
\]
The set $X$ is a domain of the IPD vector $\fat{p} =
\fat{p}_1\times\cdots\times\fat{p}_N$ of the $N$-user MAC. Remark that $X$
is not convex as a whole but each $X_k$ is convex. Also we can set a
Kronecker product of faces $F_1,\cdots,F_N$ as $F\equiv F_1\times \cdots
\times F_N$ which is a sub-domain of $X$. Obviously $F$ is not convex as a
whole even each $F_k$ is convex.

{\it A pair} $(P,X)$ is assigned to the $N$-user MAC to specify a channel
matrix $P$ and a domain $X$. Here $P$ has columns
$(P(0|i_1,\cdots,i_N),\cdots,P(m-1|i_1,\cdots,i_N))^{\tt T}$ arranged in
the order of the components of $\fat{p}_1\times\cdots\times\fat{p}_N$. An
MAC is denoted in more detail by $N$-user $(n_1,\cdots,n_N;m)$-MAC
$(P,X)$.

{\it The mutual information} of the $N$-user $(n_1,\cdots,n_N;m)$-MAC
$(P,X)$ is defined by
\begin{eqnarray}
\lefteqn{I(\fat{p}_1\times\cdots\times\fat{p}_N)} \nonumber \\
&=&\displaystyle{\sum_{j,i_1,\cdots,i_N}p_1(i_1)\cdots p_N(i_N)
P(j|i_1,\cdots,i_N)} \nonumber \\
&&\cdot\log\displaystyle{\frac{P(j|i_1,\cdots,i_N)}{q(j)}}
\label{EQ:M_INFOM}
\end{eqnarray}
where $q(j)\equiv\sum_{i_1,\cdots,i_N}p_1(i_1)\cdots p_T(i_N)
P(j|i_1,\cdots,i_N)$ is an output probability of the $j$th symbol of $B$
and $\log$ is the natural logarithm.
For any $\fat{p}',\fat{p}''\in X$, a convex-linear combination
$\lambda\fat{p}'+(1-\lambda)\fat{p}''$, $0\leq\lambda\leq 1$, does not
always belong to $X$, since $X$ is not convex except $N=1$.
Therefore, $P$ is considered in general as
\emph{a non-linear} mapping from ${\fat{p}\in X}$ to $\fat{q}\equiv
(q(0),\cdots,q(m-1))^{\tt T}$: $\fat{q}=P{\fat{p}}=$
$P(\fat{p}_1\times\cdots\times\fat{p}_N)$. Also
$I(\fat{p}_1\times\cdots\times\fat{p}_N)$ is regarded as a
multi-variables function defined on the domain
$X=X_1\times\cdots\times X_N$ and is concave (convex-above) on each $X_k$,
when $\fat{p}_\ell$'s, $\ell\neq k$, are fixed, but is not concave on the
whole domain $X$.

{\it The channel capacity} of the $N$-user $(n_1,\cdots,n_N;m)$-MAC
$(P,X)$ is defined as usual by the maximum value of the mutual information
(\ref{EQ:M_INFOM}):
\begin{equation}
C =\max_{\sfat{p}_1\times\cdots\times\sfat{p}_N\in X}
I(\fat{p}_1\times\cdots\times\fat{p}_N).
\label{EQ:CC}
\end{equation}
An IPD vector which achieves the channel capacity is referred to as
{\it an optimal} IPD vector.

{\it The Kuhn-Tucker conditions} are introduced as the conditions to
obtain the local extrema of a function of several variables subject
to one or more constraints. For the mutual information
(\ref{EQ:M_INFOM}) of the $N$-user $(n_1,\cdots,n_N;m)$-MAC $(P,X)$,
the conditions to take the maximum value (channel capacity) are
stated as follows: If $\fat{p}_1\times\cdots\times\fat{p}_N$ is
optimal, then it satisfies
\begin{equation}
\begin{array}{l}
J(\fat{p}_1 \times \cdots \times \fat{p}_N;i_k) \left\{
\begin{array}{cl}
= C, & p_k(i_k)>0 \\
\leq C, & p_k(i_k) =0
\end{array} \right. \\
\ \ i_k=0,\cdots, n_k-1,\ \ \ \ k=1,\cdots, N\\
\ \ C=I(\fat{p}_1\times\cdots\times\fat{p}_N)
\end{array}
\label{EQ:MAC-KT}
\end{equation}
where
\[
J(\fat{p}_1 \times\cdots\times \fat{p}_N;i_k)\equiv \frac{\partial
I(\fat{p}_1\times\cdots \times\fat{p}_N)} {\partial p_k(i_k)}+1
\]
\begin{eqnarray*}
\lefteqn{=\sum_{j,i_1,\cdots,i_{k-1},i_{k+1}\cdots,i_N}
p_1(i_1)\cdots p_{k-1}(i_{k-1})p_{k+1}(i_{k+1})}\\
&&\cdots p_N(i_N) P(j|i_1,\cdots,i_N)\log
\frac{P(j|i_1,\cdots,i_N)}{q(j)}.
\end{eqnarray*}
These equations (\ref{EQ:MAC-KT}) are collectively referred to as {\it the
Kuhn-Tucker conditions} for the mutual information (\ref{EQ:M_INFOM}).
These are quite easy to obtain, for example, by a method of Lagrange
multipliers to maximize the mutual information (\ref{EQ:M_INFOM}) subject
to the constraints of $\fat{p}_k$: $\sum_{i_k=0}^{n_k-1}p_k(i_k)=1$,
$k=1,\cdots,N$. Remark that the Kuhn-Tucker conditions (\ref{EQ:MAC-KT})
are obviously necessary but not in general sufficient for the channel
capacity of the MAC $(P,X)$. In the case of DMC, however, they are
necessary and sufficient for the channel capacity~\cite{Gallager}.

{\it A sub-MAC $(P,Y)$}, or a sub-channel, of an $N$-user
$(n_1,\cdots,n_N;m)$-MAC $(P,X)$ is reasonably defined as an
$N$-user MAC where the channel matrix is set to the same $P$ and the
domain is assigned to a non-empty subset $Y$ of $X$. In the
subsequent discussions we focus mostly on the sub-MAC $(P,Y)$ where
$Y$ is restricted to a sub-domain $F\equiv F_1\times\cdots\times F_N
\subseteq X_k$. Here if $\fat{p}$ is an IPD vector of the sub-MAC
$(P,F)$, then each $\fat{p}_k$ of $\fat{p}$ acts as an
$(f_k-1)$-dimensional vector on $F_k$ (i.e., $p_k(i_k)=0$ for
$i_k\not\in\Lambda({F_k})$) even it is still an
$(n_k-1)$-dimensional vector on the whole domain $X_k$ as mentioned
before. The mutual information of an $N$-user sub-MAC $(P,F)$ is
given by $I(\fat{p}\in F)$ where the $(i_1,\cdots,i_k,\cdots,i_N)$th
columns of $P$ for $i_k\not\in\Lambda({F_k})$ do not affect the
mutual information (\ref{EQ:M_INFOM}). The Kuhn-Tucker conditions
for an $N$-user sub-MAC $(P,F)$ are also given by the expression
(\ref{EQ:MAC-KT}) where $\fat{p}\in F$.

{\it The elementary MAC} now we define in general as follows: If $N$-user
$(n_1,\cdots,n_N;m)$-MAC $(P,X)$ satisfies $n_k\leq m$ for all $k =
1,\cdots,N$, then it is referred to as {\it an elementary} MAC. The
elementary MAC is an MAC whose sizes of input alphabets are not greater
than the size of output alphabet.

{\it The elementary (face) set} $\Phi_N^{(m)}$ of $X$ is defined by the
set of faces as follows: If $n_k\geq m$, then $F_k$ is put to an
$(m-1)$-dimensional face of $X_k$, and if $n_k<m$, then $F_k$ is put to
the $(n_k-1)$-dimensional $X_k$ itself. Thus the dimension of each $F_k$
is less than or equal to $(m-1)$. If $X$ is formed by $n_k\leq m$ for all
$k=1,\cdots,N$, then $\Phi_N^{(m)} =\{X\}$.

{\it A master (elementary) MAC} $(P,F)$ of an $N$-user MAC $(P,X)$
is defined as the MAC with a domain $F\in \Phi_N^{(m)}$. Here each
$\fat{p}_k\in F_k$ acts as an $(f_k-1)$-dimensional vector as
mentioned above, where $p_k(i_k)=0$ for $i_k\not\in\Lambda(F_k)$,
$k=1,\cdots,N$. Note that the master MAC $(P,F\in\Phi_N^{(m)})$ is
regarded as \emph{the largest} possible elementary MAC of $(P,X)$ in
the sense that there is no elementary MAC $(P,F')$ such that $(P,F)$
is an elementary sub-MAC of $(P,F')$. A set of all master MAC's is
referred to as the master (elementary) set of the $N$-user MAC
$(P,X)$ and is denoted by $\Omega_N$. Obviously $\Omega_N$ is finite
and is uniquely determined. If an MAC $(P,X)$ is itself elementary,
then $\Omega_N=\{\mbox{MAC} (P,X)\}$. The channel capacity of an MAC
$(P,F)\in\Omega_N$ is denoted by $C(F)$.

In later discussions we investigate the IPD vector $\fat{p}$ which
satisfies the Kuhn-Tucker conditions (\ref{EQ:MAC-KT}). If
$p_k(i_k)>0$ for all $i_k=0,\cdots,n_k-1$, $k=1,\cdots,N$, then
$\fat{p}$ is located exactly inside (not on the boundary of) $X$. If
$p_k(i_k)=0$ for $i_k\not\in\Lambda({F_k})$ and $p_k(i_k)>0$ for
$i_k\in\Lambda({F_k})$, $k=1,\cdots,N$, then the sub-domain
$F=F_1\times\cdots\times F_N$ of $X$ formed by $\Lambda({F_k})$ is
the minimum domain which contains $\fat{p}$ exactly inside $F$.

More importantly, the non-elementary MAC has in essence a
\emph{degenerate} property as follows: if $n_k>m$ for an $N$-user
$(n_1,\cdots,n_N;m)$-MAC, then for a fixed $\fat{p}_k\in
F_k\subseteq X_k$ with $f_k>m$, there exists an IPD vector
$\fat{p}_k'\in F_k'$ where $\fat{p}'\neq\fat{p}$,
$F_k'\subset(\neq)F_k$, $f_k'=m$, and
$\fat{p}=\fat{p}_1\times\cdots\times\fat{p}_k\times\cdots\times\fat{p}_N$,
$\fat{p}'=\fat{p}_1\times\cdots\times\fat{p}_k'\times\cdots\times\fat{p}_N$,
such that $\fat{q}=P\fat{p}'=P\fat{p}$. The elementary MAC has in
general no such property. This notion is crucial to the subsequent
discussions.

Finally for this section, we remark that we are going to investigate
various types of MAC's. For example, we examine an MAC $(P,Y)$ with a
domain $Y=Y_1\times\cdots\times Y_N\subset X$ where each $Y_k$,
$k=1,\cdots N$, is formed by \emph{the line} segment of IPD vectors of
$X_k$. Even then we can examine the Kuhn-Tucker conditions in the same way
as mentioned above.

\section{Main Result} \label{S:Main}
The master elementary set $\Omega_N$ as defined above has an intrinsic
property with respect to the $N$-user $(n_1,\cdots,n_N;m)$-MAC $(P,X)$. We
can state it as a main theorem:

\begin{theorem} \label{T:C(F)}
  The channel capacity $C$ of an $N$-user $(n_1,\cdots,n_N;m)$-MAC
  $(P,X)$ is achieved by the channel capacity $C(F)$ of an $N$-user elementary
  MAC $(P,F\in \Phi_N^{(m)})$ of $\Omega_N$ as follows:
\begin{equation}
C=\max_{F\in \Phi_N^{(m)}}C(F) \label{EQ:C(F)}.
\end{equation}
\end{theorem}

\proof It is sufficient to prove the case that the original MAC $(P,X)$ is
not elementary. Let $\bar{\fat{p}}=$
$\bar{\fat{p}}_1\times\cdots\times\bar{\fat{p}}_k\times\cdots\times\bar{\fat{p}}_N$
be an optimal IPD vector that achieves the channel capacity $C$. Let
$\bar{F}_k$ be the minimum face of $X_k$ which contains $\bar{\fat{p}}_k$
exactly inside $\bar{F}_k$, $k=1,\cdots N$. It is sufficient to assume
that $\bar{F}_k$ is the $m$ or more dimensional face. Then by the
degenerate property there exists an $(m-1)$-dimensional face $\tilde{F}_k
\subset(\neq)\bar{F}_k$ such that for an IPD vector
$\tilde{\fat{p}}_k\in\tilde{F}_k$,
\begin{equation}
P\bar{\fat{p}}=P(\bar{\fat{p}}_1\times\cdots\times\tilde{\fat{p}}_k
\times\cdots\times\bar{\fat{p}}_N). \label{EQ:bar(q)}
\end{equation}
Put
\[
K(\theta)\equiv I(\bar{\fat{p}}_1\times\cdots\times
(\theta\bar{\fat{p}}_k+(1-\theta)
\tilde{\fat{p}}_k)\times\cdots\times\bar{\fat{p}}_N)
\]
for the mutual information of the original MAC $(P,X)$, where
$0\leq\theta\leq 1$. The derivative $\partial{}K(\theta)/\partial\theta$
is constant by (\ref{EQ:bar(q)}) and moreover
$\partial{}K(\theta)/\partial\theta$ is equal to zero since
$\bar{\fat{p}}$ is optimal. Then it holds
\[
I(\bar{\fat{p}})=I(\bar{\fat{p}}_1\times\cdots\times\tilde{\fat{p}}_k
\times\cdots\times\bar{\fat{p}}_N).
\]
This implies that the optimal IPD vector exists in a domain
$F=F_1\times\cdots\times \tilde{F_k}\times\cdots\times
F_N\in\Phi_N^{(m)}$. Thus Theorem~\ref{T:C(F)} is proved.
\endproof

Theorem~\ref{T:C(F)} states that the channel capacity $C$ of any $N$-user
$(n_1,\cdots,n_N;m)$-MAC $(P,X)$ is rigorously determined by the channel
capacity $C(F)$ of an $N$-user master elementary MAC $(P,F)\in\Omega_N$.
In other words, an optimal IPD vector exists at least on a domain $F\in
\Phi_N^{(m)}$. However Theorem~\ref{T:C(F)} does not guarantee that the
optimal IPD vector exists only on a domain $F\in\Phi_N^{(m)}$, that is,
there \emph{might} exist in general an optimal IPD vector that is located
exactly inside $X$ and not on any $F\in\Phi_N^{(m)}$. Note that if the
$N$-user $(n_1,\cdots,n_N;m)$-MAC $(P,X)$ is elementary, then
Theorem~\ref{T:C(F)} appears self-evident since $\Omega_N$ contains only
an MAC $(P,X)$ itself.

In the remaining section of this paper we focus on the proof that the
Kuhn-Tucker conditions of an elementary MAC $(P,X)$ are necessary and
sufficient for the channel capacity. We will state it in advance as a
second theorem.

\begin{theorem} \label{T:E-MAC}
  The Kuhn-Tucker conditions for the channel capacity
  $C$ of an $N$-user elementary $(n_1,\cdots,n_N;m)$-MAC $(P,X)$, where
  $n_k\leq m$ for all $k=1,\cdots,N$, are necessary and sufficient.
\end{theorem}

It is sufficient to prove only the sufficiency since the necessity is
self-evident. From these two theorems the MAC in general can be regarded
as simply an aggregate of a finite number of elementary MAC's where the
Kuhn-Tucker conditions for the channel capacity are necessary and
sufficient.

\section{Features of Elementary MAC} \label{S:BEQ}
In this section we prepare basic properties that are required to prove the
sufficiency of Theorem~\ref{T:E-MAC}.

The first property \emph{A} is  \emph{the chain rules}~\cite{ELEMENTS}: We
recall that the mutual information of an $N$-user MAC is in general
decomposed into $N$ components with $N!$ different decompositions by the
chain rules.

The second property \emph{B} is \emph{the capacity region}: We describe
that the capacity region of the $N$-user MAC is given by the
convex-closure of all achievable rate regions of the $N!$ decompositions
for the mutual information~\cite{Ahlswede}. It is summarized as
Proposition~\ref{P:Ahlswede}.

The third property \emph{C} is \emph{the boundary equations}: We
investigate that a boundary of an achievable rate region satisfies by a
method of Lagrange multipliers a set of conditions to be referred to as
the boundary equations for the capacity region of the $N$-user MAC.

The fourth property \emph{D} is \emph{a relation between the Kuhn-Tucker
equations and the boundary equations}: We prove as Proposition~\ref
{P:KT=BEQ} that a solution of the Kuhn-Tucker conditions of an $N$-user
MAC with some restrictions satisfies the boundary equations.

The fifth property \emph{E} is \emph{local maximum}: We prove as
Proposition~\ref {P:MAXIMAL} that every solution of the Kuhn-Tucker
conditions of an elementary MAC $(P,X)$ is local maximum in the domain
$X$. To prove Proposition~\ref {P:MAXIMAL} we need Proposition~\ref
{P:KT=BEQ}.

Finally, the sixth property \emph{F} is \emph{connectedness}: We prove as
Proposition~\ref {P:CONNECTED} that a set of IPD vectors of an elementary
MAC $(P,X)$, for which the value of the mutual information is not smaller
than the arbitrary positive number, is connected in the domain $X$. To
prove Proposition~\ref {P:CONNECTED} we use Proposition~\ref {P:MAXIMAL}.

We emphasize here that the last two properties, i.e. \emph{local maximum}
and \emph{connectedness}, are the most distinctive features exclusive to
the elementary MAC. However the first four properties, although they hold
for any MAC in general, are required to step by step prove the last two.

\subsection{Chain Rules}\label{SS:ChainRule}
The mutual information of an $N$-user $(n_1,\cdots,n_N;m)$-MAC $(P,X)$ is
decomposed into $N$ components by the chain rules~\cite{ELEMENTS}. For the
IPD vectors $\fat{p}_1,\cdots,\fat{p}_{k-1}$, $\fat{p}_{k}$,
$\fat{p}_{k+1},\cdots,\fat{p}_N$, let $\fat{\rho}_{\{u,\cdots, w\}}$ be a
Kronecker product of $\fat{p}_k$, $k \not\in \{u,\cdots, w\}$, and let
$\fat{\sigma}_{\{u,\cdots,w\}}$ be a Kronecker product of $\fat{p}_k$,
$k\in\{u,\cdots,w\}$. Obviously $\fat{\sigma}_{\{u\}}=\fat{p}_u$.

The mutual information (\ref{EQ:M_INFOM}) is decomposed into two
components as
\begin{eqnarray*}
\lefteqn{ I(\fat{p}_1\times \cdots \times \fat{p}_N)=}
\nonumber \\
&&I(\fat{\sigma}_{\{u\}}|\fat{\rho}_{\{u\}})+
    I(\fat{\rho}_{\{u\}}/\fat{\sigma}_{\{u\}}).
\end{eqnarray*}
Here
\begin{eqnarray*}
\lefteqn{I(\fat{\sigma}_{\{u\}}|\fat{\rho}_{\{u\}})=} \\
&&\sum_{j,i_1,\cdots,i_N}p_1(i_1)\cdots p_N(i_N)P(j|i_1,\cdots,i_N) \\
&&\cdot\log\frac{P(j|i_1,\cdots,i_N)}
  {\sum_h p_u(h)P(j|i_1,\cdots,h,\cdots,i_N)}
\end{eqnarray*}
which is the conditional mutual information of $\fat{p}_u$ with respect to
$\fat{p}_1,\cdots,\fat{p}_{u-1},\fat{p}_{u+1}\cdots,\fat{p}_N$, and
\begin{eqnarray*}
\lefteqn{I(\fat{\rho}_{\{u\}}/\fat{\sigma}_{\{u\}})=} \\
&&\sum_{j,i_1,\cdots,i_N}p_1(i_1)\cdots p_N(i_N)P(j|i_1,\cdots,i_N) \\
&&\cdot\log\frac{\sum_h p_u(h)P(j|i_1,\cdots,h,\cdots,i_N)}
   {q(j)}
\end{eqnarray*}
which is the mutual information of an $(N-1)$-user MAC with the channel
matrix $[\sum_hp_u(h)P(j|i_1,\cdots,h,\cdots,i_N)]$. Moreover we decompose
the latter into
\begin{eqnarray*} \lefteqn{
I(\fat{\rho}_{\{u\}}/\fat{\sigma}_{\{u\}})=} \nonumber \\
&&I(\fat{p}_w|\fat{\rho}_{\{u,w\}}/\fat{\sigma}_{\{u\}})+
I(\fat{\rho}_{\{u,w\}}/\fat{\sigma}_{\{u,w\}}).
\end{eqnarray*}
In general,
\begin{eqnarray*}
\lefteqn{I(\fat{\rho}_{\{u,\cdots,w\}}/\fat{\sigma}_{\{u,\cdots,w\}})=}
\nonumber \\
&&I(\fat{p}_x|\fat{\rho}_{\{u,\cdots,x,\cdots,w\}}
/\fat{\sigma}_{\{u, \cdots, w\}}) \nonumber \\
&&+I(\fat{\rho}_{\{u,\cdots,x,\cdots,w\}}/
\fat{\sigma}_{\{u,\cdots,x\cdots,w\}}).
\end{eqnarray*}
Here
\begin{eqnarray*}
\lefteqn{ I(\fat{p}_x|\fat{\rho}_{\{u,\cdots,x,\cdots,w\}}
/\fat{\sigma}_{\{u,\cdots, w\}})=} \\
&&\sum_{j,i_1,\cdots,i_N}p_1(i_1)\cdots p_T(i_N)P(j|i_1,\cdots,i_N) \\
&&\cdot\log \frac{\langle\fat{\sigma}_{\{u,\cdots, w\}}\cdot P\rangle}
{\langle\fat{\sigma}_{\{u,\cdots,x,\cdots, w\}}\cdot P\rangle}
\end{eqnarray*}
\begin{eqnarray*}
\lefteqn{ I(\fat{\rho}_{\{u,\cdots,x,\cdots,w\}}
/\fat{\sigma}_{\{u, \cdots,x \cdots, w\}})=} \\
&&\sum_{j,i_1,\cdots,i_N}p_1(i_1)\cdots p_N(i_N)P(j|i_1,\cdots,i_N) \\
&& \cdot\log\frac{\langle\fat{\sigma}_{\{u,\cdots,x,\cdots, w\}}
  \cdot P\rangle}{q(j)}
\end{eqnarray*}
where
\begin{eqnarray*}
\lefteqn{\langle\fat{\sigma}_{\{u,\cdots,w\}}\cdot P\rangle\equiv
\sum_{h_u,\cdots,h_w}p_u(h_u)\cdots p_w(h_w)} \\
&&\cdot P(j|i_1,\cdots,h_u,\cdots,h_w,\cdots,i_N).
\end{eqnarray*}
If $\{u,\cdots,w\}$ is empty, then $\langle\fat{\sigma}_{\{u,\cdots,
  w\}}\cdot{P}\rangle$ reduces to $P(j|i_1,\cdots,i_N)$. Thus successively
reducing the suffices $\{u,\cdots,w\}$ of $\fat{\rho}_{\{u,\cdots,w\}}$ up
to $\{1,\cdots,k-1,k+1,\cdots,N\}$,
$I(\fat{p}_1\times\cdots\times\fat{p}_N)$ is decomposed into $N$
components. Note that there exist as a whole $N!$ different
decompositions.

\subsection{Capacity Region}\label{SS:CapacityRegion}
A set of all achievable rates for an $N$-user $(n_1,\cdots,n_N;m)$-MAC
$(P,X)$ is called a capacity region (e.g., \cite{ELEMENTS,Liao,Ahlswede}).
By a decomposition we obtain
\begin{eqnarray}
\lefteqn{
I(\fat{p}_1\times \cdots\times \fat{p}_N)=} \nonumber \\
&&I(\fat{p}_1|\fat{p}_2\times \cdots\times \fat{p}_N)+
 I(\fat{p}_2|\fat{p}_3\times
\cdots\times \fat{p}_N/\fat{p}_1)
\nonumber \\
&&+I(\fat{p}_3|\fat{p}_4\times \cdots\times \fat{p}_N
/\fat{p}_1\times\fat{p}_2)+\cdots \nonumber \\
&&\cdots+ I(\fat{p}_N/\fat{p}_1\times\cdots\times\fat{p}_{N-1}).
 \label{EQ:D_1-T}
\end{eqnarray}
There exist as a whole $N!$ different decompositions as mentioned above.
Define a sub-region $G_1$ as
\begin{eqnarray*}
  \bigcup_{\sfat{p}\in X}
(I(\fat{p}_1|\fat{p}_2\times\cdots\times\fat{p}_N),\cdots,
I(\fat{p}_N/\fat{p}_1\times\cdots\times\fat{p}_{N-1})).
\end{eqnarray*}
This is identified as a set of achievable rates $G_1$ for the
decomposition (\ref{EQ:D_1-T}). Other $N!-1$ sets of achievable rates
$G_2$, $\cdots$, $G_{N!}$ are also defined in the same way as $G_1$. Then
the capacity region $G$ is determined by those sub-regions $G_i$'s as
follows [Theorem~15.3.6 in \cite{ELEMENTS}]:
\begin{proposition}\label{P:Ahlswede}
The capacity region of an $N$-user $(n_1,\cdots,n_N;m)$-MAC $(P,X)$ is
given by
\begin{equation}
G=\mbox{\rm co}\bigcup_{i=1}^{N!} G_i
\end{equation}
where ``{\rm co}'' implies the convex-closure.
\endtheorem
\end{proposition}

\subsection{Boundary Equations}\label{SS:BoundaryEquation}
\begin{figure}[t]
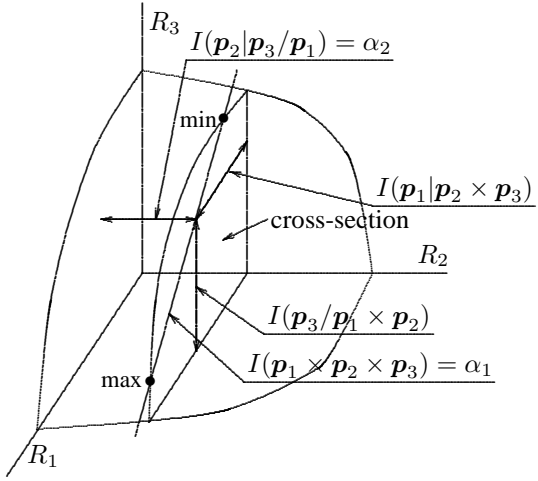

\centerline{%
\beginpicture
\setcoordinatesystem units <.9mm,.9mm> point at 15 5 \setplotarea x from
-20 to 66, y from -30 to 40
\plot -20 -30 0 0 / \put {$R_1$} [l] at -17 -27 \plot 45 0 0 0 / \put
{$R_2$} [b] at 43 1 \plot 0 40 0 0 / \put {$R_3$} [l] at 1 37
\setquadratic \plot -15.5 -23 -14.3 -10 -13.5 -4 / \plot -8 16 -5.2 22 0
30 /
\plot -13.5 -4 -11.2 6.1 -8 16 /
\setquadratic \plot -15.5 -23 4 -21.5 13 -20 / \plot 26 -14 28.8 -10.2 34
0 /
\plot 13 -20 19.4 -17.5 26 -14 /
\setquadratic \plot 34 0 31.9 12.7 30 18 / \plot 24 24 17 26.7 0 30 /
\plot 24 24 27.3 21.3 30 18 /
\setquadratic \plot 15.5 27 10 20 6.8 14.1 / \plot 1 -22 1.5 -10 2 -4 /
\plot 2 -4 3.7 5 6.8 14.1 /
%
\setlinear \plot -1 -24 14 30 / \plot 1 -22 15.5 0 / \plot 15.5 27 15.5 0
/
\multiput {\small $\bullet$} at 1.2 -16 12 22.8 / \put {\small\rm min} [r]
at 11 22.8 \put {\small\rm max} [r] at 0 -16 \plot 15 -16 52 -16 / \arrow
<4pt> [0.2,0.5] from 15 -16 to 4 -6 \put
{$I(\fat{p}_1\times\fat{p}_2\times\fat{p}_3)=\alpha_1$}
 [bl] at 15.5 -15.5
\arrow <4pt> [0.2,0.5] from 8 8 to 8 -11.5 \arrow <4pt> [0.2,0.5] from 8
-11.5 to 8 8 \plot 43 -9 18 -9 / \arrow <4pt> [0.2,0.5] from 18 -9 to 8 -2
\put {$I(\fat{p}_3/\fat{p}_1\times\fat{p}_2)$} [bl] at 18.5 -8.5
\put {cross-section} [l] at 19 8 \arrow <4pt> [0.2,0.5] from 18.5 8 to 12
5
\arrow <4pt> [0.2,0.5] from 8 8 to 15.5 19.5 %
\arrow <4pt> [0.2,0.5] from 15.5 19.5 to 8 8 \put
{$I(\fat{p}_1|\fat{p}_2\times\fat{p}_3)$} [lb] at 34.5 10.5 \plot 58 10 34
10 / \arrow <4pt> [0.2,0.5] from 34 10 to 13 15
\arrow <4pt> [0.2,0.5] from 8 8 to -6 8 \arrow <4pt> [0.2,0.5] from -6 8
to 8 8 \put {$I(\fat{p}_2|\fat{p}_3/\fat{p}_1)=\alpha_2$} [lb] at 7 32
\plot 6.5 31.5 37 31.5 / \arrow <4pt> [0.2,0.4] from 6.5 31.5 to 2 8
\endpicture
} \caption{Sub-region $G_1$ of three-user MAC.} \label{F:3-U_CAP}
\end{figure}

A boundary of each sub-region $G_i$, $i$ = $1,\cdots,N!$, for an $N$-user
$(n_1,\cdots,n_N;m)$-MAC $(P,X)$, can be determined by a method of
Lagrange multipliers. The boundary of $G_1$, for example, is evaluated by
a Lagrange multiplier function,
\begin{eqnarray*}
\lefteqn{L(\fat{p}_1,\cdots,\fat{p}_N,\lambda_1,\cdots,\lambda_{N-1},
\zeta_1,\cdots,\zeta_N)=} \\
&& I(\fat{p}_1|\fat{p}_2\times\cdots\times\fat{p}_N) \\
&& -\lambda_1I(\fat{p}_1\times\cdots\times\fat{p}_N) \\
&& -\lambda_2I(\fat{p}_2|\fat{p}_3\times\cdots\times\fat{p}_N/ \fat{p}_1)
- \cdots \\
&&-\lambda_{N-1}I(\fat{p}_{N-1}|\fat{p}_N/\fat{p}_1\times\cdots
\times\fat{p}_{N-2}) \\
&& - \sum_{k=1}^N\zeta_k\sum_{i_k}p_k(i_k)
\end{eqnarray*}
where $\lambda_1,\cdots,\lambda_{N-1}$ and $\zeta_1,\cdots,\zeta_N$ are
so-called Lagrange multipliers. The conditions that an IPD vector
$\fat{p}_1 \times \cdots \times \fat{p}_N$ takes extremum (maximum or
minimum) for $G_1$ are given by the equations (see Fig.~\ref{F:3-U_CAP}
for $N$ = 3)
\[
\det\left[
\begin{array}{cc}
\displaystyle{ \frac{\tilde{\partial}I(\fat{p}_1|\fat{p}_2\times
\cdots\times\fat{p}_N)} {\tilde{\partial}p_1(i_1)}}& \displaystyle{
\frac{\tilde{\partial}I(\fat{p}_1\times\cdots\times\fat{p}_N)}
{\tilde{\partial}p_1(i_1)}} \\
\vdots & \vdots \\
\displaystyle{ \frac{\tilde{\partial}I(\fat{p}_1|\fat{p}_2\times
\cdots\times\fat{p}_N)} {\tilde{\partial}p_N(i_N)}}& \displaystyle{
\frac{\tilde{\partial}I(\fat{p}_1\times\cdots\times\fat{p}_N)}
{\tilde{\partial}p_N(i_N)}} \\
\end{array}\right.
\]
\[
\begin{array}{cc}
\displaystyle{\frac{\tilde{\partial}I(\fat{p}_2|\fat{p}_3\times
\cdots\times\fat{p}_N/
\fat{p}_1)}{\tilde{\partial}p_1(i_1)}}& \cdots \\
\vdots & \vdots \\
\displaystyle{\frac{\tilde{\partial}I(\fat{p}_2|\fat{p}_3\times
\cdots\times\fat{p}_N/\fat{p}_1)}{\tilde{\partial}p_N(i_N)}}&\cdots
\end{array}
\]
\[
\left.
\begin{array}{cc}
\cdots& \displaystyle{
\frac{\tilde{\partial}I(\fat{p}_{N-1}|\fat{p}_N/\fat{p}_1\times\cdots\times
\fat{p}_{N-2})}
{\tilde{\partial}p_1(i_1)}} \\
\vdots & \vdots \\
\cdots& \displaystyle{
\frac{\tilde{\partial}I(\fat{p}_{N-1}|\fat{p}_N/\fat{p}_1\times\cdots\times
\fat{p}_{N-2})} {\tilde{\partial}p_N(i_N)}}
\end{array}\right]
\]
\begin{equation}
=0, \ \ i_k=0,\cdots,n_k-2, k=1,\cdots,N. \label{EQ:TU_BOUNDARY}
\end{equation}
Here, we define partial derivatives as:
\[
\frac{\tilde{\partial}I(\cdots)}{\tilde{\partial}p_k(i_k)}\equiv
 \frac{\partial I(\cdots)}{\partial p_k(i_k)}- \frac{\partial I(\cdots)}
  {\partial p_k(n_k-1)}, i_k\neq n_k-1.
\]
Total $(n_1-1)\times\cdots\times(n_N-1)$ equations (\ref{EQ:TU_BOUNDARY})
are collectively referred to as {\it the boundary equations} for $G_1$.
Solutions of (\ref{EQ:TU_BOUNDARY}) include both maximization and
minimization as usual. Successively we can set up the boundary equations
for $G_2,\cdots,G_{N!}$ with totally the same form as
(\ref{EQ:TU_BOUNDARY}). Note that the boundary equations have the same
form as (\ref{EQ:TU_BOUNDARY}) for the different choices of starting
Lagrange multiplier function.

\subsection{A relation between the Kuhn-Tucker equations and the boundary
 equations} \label{SS:KTC}
The boundary equations thus obtained have an
important property which we state as a proposition:
\begin{proposition} \label{P:KT=BEQ}
If a solution
$\bar{\fat{p}}=\bar{\fat{p}}_1\times\cdots\times\bar{\fat{p}}_N\in X$ of
the Kuhn-Tucker conditions for the mutual information $I(\fat{p})$ of an
$N$-user $(n_1,\cdots,n_N;m)$-MAC $(P,X)$ satisfies
\begin{eqnarray}
\lefteqn{
J(\bar{\fat{p}}_1\times\cdots\times\bar{\fat{p}}_N;i_k)=C,} \nonumber \\
&& i_k=0,\cdots,n_k-1, k=1,\cdots,N \nonumber \\
&& C=I(\bar{\fat{p}}_1\times\cdots\times\bar{\fat{p}}_N)
 \label{EQ:J=C}
\end{eqnarray}
then $\bar{\fat{p}}$ is a solution of the boundary equations for
sub-regions $G_i$, $i=1,\cdots, N!$.  \endtheorem
\end{proposition}

\proof It is sufficient to prove that $\bar{\fat{p}}$ satisfies the
boundary equation (\ref{EQ:TU_BOUNDARY}) for $G_1$. By the assumption
(\ref{EQ:J=C}), it holds
\begin{eqnarray*}
\lefteqn{\left.\frac{\tilde{\partial}I(\fat{p}_1\times\cdots\times\fat{p}_N)}
{\tilde{\partial}p_k(i_k)}
\right|_{\sfat{p}_k=\bar{\sfat{p}}_k,i_k=0,\cdots,n_k-2}=} \\
&&J(\bar{\fat{p}}_1 \times \cdots \times \bar{\fat{p}}_N;i_k) -
J(\bar{\fat{p}}_1 \times \cdots \times \bar{\fat{p}}_N;n_k-1) \\
&&=0.
\end{eqnarray*}
Then the second column of (\ref{EQ:TU_BOUNDARY}) reduces to zeros.
Therefore $\bar{\fat{p}}$ is a solution of the boundary equation
(\ref{EQ:TU_BOUNDARY}).
\endproof

Remark that Proposition~\ref{P:KT=BEQ} holds for any MAC including the
elementary MAC if it satisfies the conditions (\ref{EQ:J=C}).

\subsection{Local Maximum}\label{SS:LocalMaximum}
An IPD vector $\bar{\fat{p}}$ is called {\it a local maximum point} for
the mutual information $I(\fat{p})$, if there exists a neighborhood
$U_{\bar{\sfat{p}}}$ of $\bar{\fat{p}}$ such that $I(\fat{p})\leq
I(\bar{\fat{p}})$ for any $\fat{p}\in U_{\bar{\sfat{p}}}$. We prove here
that for \emph{the elementary} MAC every solution of the Kuhn-Tucker
conditions is local maximum. We state it as a proposition:
\begin{proposition} \label{P:MAXIMAL}
  If an $N$-user $(n_1,\cdots,n_N;m)$-MAC $(P,X)$ is elementary, i.e. $n_k\leq
  m$, $k=1,\cdots,N$, then every solution
  $\fat{p}^*\equiv\fat{p}_1^*\times\cdots\times\fat{p}_N^*\in X$ of the
  Kuhn-Tucker conditions for the mutual information $I(\fat{p})$ is local
  maximum in $X$.
  \endtheorem
\end{proposition}

Before proceeding we remark that Proposition~\ref{P:MAXIMAL} does not hold
in general for the non-elementary MAC by the degenerate property as is
stated in the beginning of the proof of Theorem~\ref{T:C(F)}. In fact, we
note without proof that a non-elementary two-user $(3,3;2)$-MAC $(P,X)$,
for example, with $N=2$, $n_1=n_2=3$, $m=2$, for some channel matrix $P$,
has a solution of the Kuhn-Tucker conditions which is not local maximum in
$X$.

\proof Since the $N$-user MAC $(P,X)$ is elementary, it is sufficient to
investigate two cases for the solution $\fat{p}^*$ of the Kuhn-Tucker
conditions: the first is that every $p_k^*(i_k)$ is non-zero and the
second is that at least one of $p_k^*(i_k)$'s is zero.

In the first case, since $p_k^*(i_k)>0$ for all components, $\fat{p}^*$
satisfies the Kuhn-Tucker conditions:
\begin{eqnarray}
\lefteqn{J(\fat{p}_1^*\times\cdots\times\fat{p}_N^*;i_k)=M_T} \nonumber \\
&& p_k^*(i_k)>0, \ \ i_k=0,\cdots, n_k-1, \ \ k=1,\cdots,N \nonumber \\
&& M_T=I(\fat{p}_1^*\times\cdots\times\fat{p}_N^*). \label{EQ:KTCM}
\end{eqnarray}
and there exist $\fat{p}'$ and $\fat{p}''$ in $X$ such that
\[
\fat{p}^*= (\theta_1^*\fat{p}_1''+(1-\theta_1^*)\fat{p}_1')
\times\cdots\times (\theta_N^*\fat{p}_N''+(1-\theta_N^*)\fat{p}_N')
\]
where $\fat{p}_k^*\neq\fat{p}_k'$, $\fat{p}_k^*\neq\fat{p}_k''$, and
$0<\theta_k^*<1$.

Here we put by using $\theta_k$,  $0\leq\theta_k{\leq}1$, $k=1,\cdots, N$,
\begin{eqnarray*}
\lefteqn{
{\cal K}(\theta_1,\cdots,\theta_N)=} \\
&&I((\theta_1\fat{p}_1''+(1-\theta_1)\fat{p}_1')\times\cdots\times
(\theta_N\fat{p}_T''+(1-\theta_N)\fat{p}_N'))
\end{eqnarray*}
and investigate the Kuhn-Tucker conditions and the boundary equations with
respect to this ${\cal K}(\theta_1,\cdots,\theta_N)$. The Kuhn-Tucker
conditions are simple to see as
\begin{equation}
{\cal K}_k(\theta_1,\cdots,\theta_N)=0,\ \ k=1,\cdots,N \label{EQ:KTK}
\end{equation}
where ${\cal K}_k(\theta_1,\cdots,\theta_N)\equiv{\partial {\cal
K}(\theta_1,\cdots,\theta_N)}/{\partial\theta_k}$. Also by a decomposition
\begin{eqnarray*}
\lefteqn{{\cal K}(\theta_1,\cdots,\theta_N)=} \\
&&{\cal K}(\theta_1|\theta_2,\cdots,\theta_N)+
{\cal K}(\theta_2|\theta_3,\cdots,\theta_N/\theta_1) \\
&&+{\cal K}(\theta_3|\theta_4\cdots,\theta_N/\theta_1,\theta_2)
+\cdots \\
&& +{\cal K}(\theta_N/\theta_1,\cdots,\theta_{N-1})
\end{eqnarray*}
we obtain a set of achievable rates
\begin{equation}
{\cal G}_1=\bigcup_{\theta_1,\cdots,\theta_N} ({\cal
K}(\theta_1|\theta_2,\cdots,\theta_N),\cdots, {\cal
K}(\theta_N/\theta_1,\cdots,\theta_{N-1})) \label{EQ:ARR-G}
\end{equation}
which leads us to the boundary equation for ${\cal{G}}_1$ as follows:
\[
\det\left[
\begin{array}{cc}
{\cal K}_1(\theta_1|\theta_2,\cdots,\theta_N)&{\cal K}_1(\theta_1,
    \cdots,\theta_N)\\
  \vdots&\vdots  \\
{\cal K}_N(\theta_1|\theta_2,\cdots,\theta_N)&{\cal K}_N(\theta_1,
   \cdots,\theta_N)
\end{array}\right.
\]
\[
\begin{array}{cc}
{\cal K}_1(\theta_2|\theta_3,\cdots,\theta_N/\theta_1) & \cdots \\
  \vdots &  \vdots  \\
{\cal K}_N(\theta_2|\theta_3,\cdots,\theta_N/\theta_1) & \cdots
\end{array}
\]
\begin{equation}
\left.
\begin{array}{c}
{\cal K}_1(\theta_{N-1}|\theta_N/\theta_1,\cdots,\theta_{N-2}) \\
\vdots \\
{\cal K}_N(\theta_{N-1}|\theta_N/ \theta_1,\cdots, \theta_{N-2})
\end{array}\right]=0 \label{EQ:K-BEQ}
\end{equation}
where ${\cal K}_k(\cdots)\equiv{\partial{\cal
K(\cdots)}}/{\partial\theta_k}$.

Since $\fat{p}^*$ satisfies (\ref{EQ:KTCM}),
$(\theta_1^*,\cdots,\theta_N^*)$ is a solution of the Kuhn-Tucker
conditions (\ref{EQ:KTK}). Then by Proposition~\ref{P:KT=BEQ} it satisfies
the boundary equation (\ref{EQ:K-BEQ}) and $M_T={\cal
K}(\theta_1^*,\cdots,\theta_N^*)$.

\begin{figure}[t]
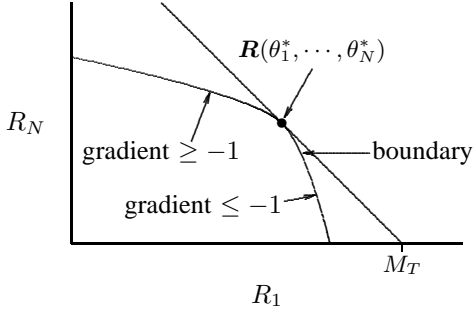

\centerline{%
\beginpicture
\setcoordinatesystem units <.8mm,.8mm>
\setplotarea x from 0 to 65, y from 0 to 40
\axis left label {$R_N$} ticks length <0pt> /
\axis bottom label {$R_1$} ticks length <3pt> withvalues {} / at 55 / /
\plot 15 40 55 0 /
\put {\small$\bullet$} at 35 20
\put {\small$M_T$} [t] at 55 -2
\put {boundary} [l] at 50 15
\arrow <4pt> [0.2,0.4] from 49.5 15 to 38.6 15
\put {\small$\fat{R}(\theta_1^*,\cdots,\theta_N^*)$} [b] at 40 30
\arrow<4pt> [0.2,0.4] from 40 29 to 35.5 21
\put {gradient $\leq-1$} [rt] at 35 8
\arrow <4pt> [.2,.4] from 35.5 7 to 40.5 8
\put {gradient $\geq-1$} [rt] at 28 17
\arrow <4pt> [.2,.4] from 20 17 to 23 25
\setquadratic
\plot 0 31  24 25  35 20 /
\plot 35 20 39 13   43 0 /
\endpicture
} \caption{Boundary of cross-section ${\cal G}_1(R_1,R_N)$ in $R_1$-$R_N$
  plain. Rates $R_2,\cdots,R_{N-1}$ are fixed as specified by
  (\ref{EQ:K-SUBJECT}).}  \label{F:G_K}
\end{figure}

Now we examine a gradient of the boundary of ${\cal G}_1$ at $\theta^*
\equiv (\theta_1^*,\cdots,\theta_N^*)$. Note that the solution $\theta
\equiv (\theta_1,\cdots,\theta_N)$ of the boundary equation
(\ref{EQ:K-BEQ}) around $\theta^*$ defines a set of achievable rates
(\ref{EQ:ARR-G}) as ${\cal G}_1(R_1,\cdots, R_N)$. Obviously ${\cal
K}(\theta^*)=M_T$. At this step we investigate a cross-section of
(\ref{EQ:ARR-G}) subject to the restrictions such that
\begin{eqnarray}
{\cal K}(\theta_2|\theta_3,\cdots,\theta_N/\theta_1)&=&
{\cal K}(\theta_2^*|\theta_3^*,\cdots,\theta_N^*/\theta_1^*)\nonumber \\
& \vdots & \nonumber \\
{\cal K}(\theta_{N-1}|\theta_N/\theta_1,\cdots,\theta_{N-2})&=& {\cal
K}(\theta_{N-1}^*|\theta_N^*/\theta_1^*,\cdots,\theta_{N-2}^*). \nonumber \\
\label{EQ:K-SUBJECT}
\end{eqnarray}
We denote a cross-section (subset) of ${\cal G}_1$ subject to
(\ref{EQ:K-SUBJECT}) as ${\cal G}_1(R_1,R_N)$. This is composed of
\begin{eqnarray*}
\lefteqn{\fat{R}(\theta_1,\cdots,\theta_N)\equiv}\\
&&({\cal K}(\theta_1|\theta_2,\cdots,\theta_N),
{\cal K}(\theta_2^*|\theta_3^*,\cdots,\theta_N^*/\theta_1^*), \cdots, \\
&&{\cal
K}(\theta_{N-1}^*|\theta_{N-1}^*/\theta_1^*,\cdots,\theta_{N-2}^*), {\cal
K}(\theta_N/\theta_1,\cdots,\theta_{N-1})).
\end{eqnarray*}
The cross-section ${\cal G}_1(R_1,R_N)$ is a region in the two-dimensional
$(R_1\mbox{-}R_N)$ plain as shown in Fig.~\ref{F:G_K}. Since it holds
\[
\begin{array}{rcl}
{\cal K}_1(\theta_2|\theta_3,\cdots,\theta_N/\theta_1)&=&0 \\
&\vdots& \\
{\cal K}_1(\theta_{N-1}|\theta_N/\theta_1,\cdots,\theta_{N-2})&=&0
\end{array}
\]
by the restrictions (\ref{EQ:K-SUBJECT}), then we have ${\cal
K}_1(\theta_1,\cdots,\theta_N) = {\cal
K}_1(\theta_1|\theta_2,\cdots,\theta_{N-1}) + {\cal
K}_1(\theta_N/\theta_1,\cdots,\theta_{N-1})$. Thus the gradient of the
boundary of ${\cal G}_1(R_1,R_N)$ appears
\begin{equation}
\frac{{\cal K}_1(\theta_N/\theta_1,\cdots,\theta_{N-1})}
 {{\cal K}_1(\theta_1|\theta_2,\cdots,\theta_N)}=-1+
\frac{{\cal K}_1(\theta_1,\cdots,\theta_N)}
 {{\cal K}_1(\theta_1|\theta_2,\cdots,\theta_N)}. \label{EQ:Gradient}
\end{equation}
The right-hand side of (\ref {EQ:Gradient}) is estimated as
\begin{equation}
-1+\frac{{\cal K}_1(\theta_1,\cdots,\theta_N)} {{\cal
K}_1(\theta_1|\theta_2,\cdots,\theta_N)}\leq(\geq)\-1
\label{EQ:leq-1}
\end{equation}
according to the maximization (minimization) conditions of
${\cal{K}}(\theta_1|\theta_2,\cdots,\theta_N)$ subject to
(\ref{EQ:K-SUBJECT}) where it holds
\[
{\cal K}_1(\theta_1,\cdots,\theta_N) {\cal
K}_1(\theta_1|\theta_2,\cdots,\theta_N)\leq(\geq)\ 0.
\]
Also the gradient of the boundaries of any cross-section ${\cal
G}_1(R_1,R_k)$ ($2\leq k\leq N-1$) is given by (\ref{EQ:leq-1}).

For any region ${\cal G}_i$ of $N!$ decompositions, the gradient of the
boundary of ${\cal G}_i$ at $\theta^*$ takes the same condition as that of
${\cal G}_1$.

Since the inequalities (\ref{EQ:leq-1}) are valid for any $\fat{p}_k'$,
$\fat{p}_k''$, $k=1,\cdots,N$, there exists a neighborhood
$U_{\sfat{p}^*}$ of $\fat{p}^*$ in $X$, such that $I(\fat{p}^*)\geq
I(\fat{p}\in U_{\sfat{p}^*})$. This means that $\fat{p}^*$ is local
maximum in $X$.

In the second case, since at least one of $p_k^*(i_k)$'s is zero,
$\fat{p}^*$ satisfies the Kuhn-Tucker conditions:
\begin{equation}
\begin{array}{l}
J(\fat{p}_1^* \times \cdots \times \fat{p}_N^*;i_k) \left\{
\begin{array}{cl}
= M_T, & i_k\in\Lambda(F_k) \\
\leq M_T, & i_k\not\in\Lambda(F_k)
\end{array} \right. \\
\ \ k=1,\cdots,N, \ \ M_T=I(\fat{p}_1^* \times \cdots \times \fat{p}_N^*)
\end{array}
\label{EQ:KTCM0}
\end{equation}
where $p_k^*(i_k)$ $>0$ for $i_k\in\Lambda(F_k)$ and $p_k^*(i_k)$ $=0$ for
$i_k\not\in\Lambda(F_k)$. Thus there exists a sub-domain
$F=F_1\times\cdots\times F_N$ such that $\fat{p}^* \in F$. This implies
that $\fat{p}^*$ is local maximum in $F$ as described in the first case
and there exists a neighborhood $U_{0\sfat{p}^*}\subset F$ such that
$I(\fat{p}^*)\geq I(\fat{p}\in U_{0\sfat{p}^*})$.

For any
$\fat{p}'=\fat{p}_1'\times\cdots\times\fat{p}_k'\times\cdots\times\fat{p}_N'$
$\in U_{0\sfat{p}^*}$, consider
$\fat{p}''=\fat{p}_1'\times\cdots\times\fat{p}_k''\times\cdots\times\fat{p}_N'$,
where $\fat{p}_k''$ $\in$ $X_k$ and $\fat{p}_k''$ $\not\in$ $F_k$. Put for
$0\leq\theta\leq1$
\[
K(\theta)= I(\fat{p}_1'\times\cdots\times(\theta\fat{p}_k''
 +(1-\theta)\fat{p}_k')\times\cdots\times\fat{p}_N').
\]
It holds ${dK(\theta)}/{d\theta} \ |_{\theta=0}\leq 0$, since $K(\theta)$
is concave, differentiable, and $\fat{p}^*$ satisfies (\ref{EQ:KTCM0}).
Therefore $K(\theta)$ is monotone non-increasing for $\theta$. Thus there
exists $\theta'>0$ such that
$I(\fat{p}_1'\times\cdots\times(\theta'\fat{p}_k''+(1-\theta')\fat{p}_k')
\times\cdots\times\fat{p}_N')<I(\fat{p}^*)$. Hence, there exists a
neighborhood $U_{\sfat{p}^*}\subset X$ of $\fat{p}^*$ such that
$I(\fat{p}^*)\geq I(\fat{p}\in U_{\sfat{p}^*})$. This means that
$\fat{p}^*$ is local maximum in $X$.

By these two cases Proposition~\ref{P:MAXIMAL} is proved.
\endproof

\subsection{Connectedness}\label{SS:Conectedness}
Finally in this section, we prove the property of connectedness for the
elementary MAC as a proposition:
\begin{proposition} \label{P:CONNECTED}
If an $N$-user $(n_1,\cdots,n_N;m)$-MAC $(P,X)$ is elementary, i.e.
$n_k\leq m$, $k=1,\cdots,N$, then the set
\begin{equation}
D(a)\equiv\{\fat{p}|I(\fat{p}\in X){\geq}a\} \label{EQ:D(a)}
\end{equation}
is connected for any $a\geq0$. \endtheorem
\end{proposition}

\proof Assume that for any $\varepsilon>0$, there exists $a_0>0$ such that
$D(a_0)$ is connected and $D(a_0+\varepsilon)$ is disconnected.

\begin{figure}[t]
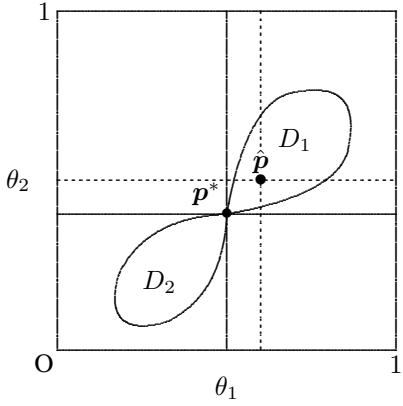

\centerline{%
\beginpicture
\setcoordinatesystem units <.9mm,.9mm> \setplotarea x from -8 to 53, y
from -8 to 53
\plot 0 0 50 0 50 50 0 50 0 0 / \plot 0 20 50 20 / \plot 25 0 25 50 / \put
{$1$} [t] at 50 -1 \put {$1$} [r] at -1 50 \put {O} at -2 -2 \put
{$\theta_1$} [t] at 25 -4 \put {$\theta_2$} [r] at -4 25 \put
{\small$\bullet$} at 25 20 \put {$\fat{p}^*$} at 22 23 \setquadratic
\plot 25 20 29 33 35 38 / \plot 35 38 42 37 43 30 / \plot 25 20 38 24 43
30 / \put {$D_1$} at 35 31
\plot 25 20 15 17.5 9 11 / \plot 25 20 22.5 10 16 4 / \plot 9 11 9.5 4.5
16 4 / \put {$D_2$} at 15 10
\setlinear \setdashpattern <.8pt, 2pt> \put {$\hat{\fat{p}}$} [b] at 30 26
\put {$\small\bullet$} at 30 25 \plot 0 25   50 25 / \plot 30 0   30 26 /
\plot 30 30  30 50 /
\endpicture
} \caption{Pattern of $D(a_0)=D_1\cup D_2$ for the case of $N=2$,
$n_1=n_2=2$, $p_1(0)=\theta_1$, $p_2(0)=\theta_2$.}\label{F:DISCONN}
\end{figure}

Since $I(\fat{p})$ is concave on each $X_k$, then there exist subsets
$D_1$ and $D_2$ of $D(a_0)$ with properties as follows:
\begin{enumerate}
\item $D(a_0)=D_1\cup D_2$, and $I(\fat{p}^*)=a_0$, for $\fat{p}^*\in
  D_1\cap D_2$.

\item For any
 $\fat{p}_1'\times\cdots\times\fat{p}_k'\times\cdots\times\fat{p}_N'\in{}D_1$,
 all IPD vectors
 $\fat{p}_1'\times\cdots\times\fat{p}_k\times\cdots\times\fat{p}_N'$,
 $\fat{p}_k\in X_k$, $k=1,\cdots,N$, satisfying
 $I(\fat{p}_1'\times\cdots\times\fat{p}_k\times\cdots\times\fat{p}_N')\geq{}a_0$
  belongs to $D_1$, and also for any
 $\fat{p}_1''\times\cdots\times\fat{p}_k''\times\cdots\times\fat{p}_N''\in{}D_2$,
 all IPD vectors
 $\fat{p}_1''\times\cdots\times\fat{p}_k\times\cdots\times\fat{p}_N''$,
 $\fat{p}_k\in X_k$, $k=1,\cdots,N$, satisfying
 $I(\fat{p}_1''\times\cdots\times\fat{p}_k\times\cdots\times\fat{p}_N'')\geq{}a_0$
 belongs to $D_2$ (cf.~Fig.~\ref{F:DISCONN}).
\end{enumerate}
%
Thus for any $\varepsilon>0$, $D(a_0+\varepsilon)$ is separated into
subsets $D_1'\subset D_1$ and $D_2'\subset D_2$ such that $D(a)=D_1'\cup
D_2'$ and $D_1'\cap D_2'=\phi$.

It is easy to see that for any
$\fat{p}^*=\fat{p}_1^*\times\cdots\times\fat{p}_k^*\times\cdots\times\fat{p}_N^*\in
D_1\cap D_2$, $k=1,\cdots N$, it holds
\begin{equation}
I(\fat{p}_1^*\times\cdots\times\fat{p}_k\times\cdots\times\fat{p}_N^*)
 \leq a_0, \ \ \fat{p}_k\in X_k \label{EQ:D_1capD_2}
\end{equation}
since for any
$\hat{\fat{p}}=\hat{\fat{p}}_1\times\cdots\times\hat{\fat{p}}_N{\in}D_1$
(or $\hat{\fat{p}}\in D_2$),
$\hat{\fat{p}}\not\in D_1\cap D_2$, every
$\hat{\fat{p}}_1\times\cdots\times\fat{p}_k\times\cdots\times\hat{\fat{p}}_N$,
$k=1,\cdots,N$, satisfying
\[
I(\hat{\fat{p}}_1\times\cdots\times\fat{p}_k\times\cdots\times\hat{\fat{p}}_N)<
a_0, \fat{p}_k\in X_k
\]
belongs to neither $D_1$ nor $D_2$ by the property 2) (see
Fig.~\ref{F:DISCONN}).

Consider two cases: Every components of $\fat{p}^*$ is non-zero and at
least a component of $\fat{p}^*$ is zero.

In the first case, it holds by (\ref{EQ:D_1capD_2}) that $\fat{p}^*\in
D_1\cap D_2$ satisfies the Kuhn-Tucker conditions
\[
\begin{array}{rcl}
J(\fat{p}_1^*\times\cdots\times\fat{p}_N^*;i_k)
&=& a_0,\ \ p_k^*(i_k)\neq 0 \\
&& i_k=0,\cdots, n_k-1 \\
&& k=1,\cdots,N.
\end{array}
\]
Therefore $\fat{p}^*$ is local maximum for $\fat{p}\in X$ by
Proposition~\ref{P:MAXIMAL} since $(n_1,\cdots,n_N;m)$-MAC $(P,X)$ is
elementary. Then there exists a neighborhood $U_{\sfat{p}^*}$ of
$\fat{p}^*$ in $X$ such that $I(\fat{p})\leq a_0$ for any $\fat{p}\in
U_{\sfat{p}^*}$.

On the other hand, by the properties of $D_1$ and $D_2$ there exists
$\fat{p}'$ in either $U_{\sfat{p}^*}\cap D_1$ or $U_{\sfat{p}^*}\cap D_2$
such that $I(\fat{p}')>a_0$. This is inconsistent with that $\fat{p}^*$ is
local maximum. Therefore $D(a)$ is connected.

For the second case, consider the minimum domain $F\equiv
F_1\times\cdots\times{F_N}\subset X$ which contains the $\fat{p}^*$
exactly inside (and not on the boundary of) $F_k$, where $p_k(i_k)=0$ for
$i_k\not\in\Lambda({F_k})$ and $p_k(i_k)>0$ for $i_k\in\Lambda({F_k})$,
$k=1,\cdots,N$. In the same way as in the first case, it is proved that
$D(a)\cap F$ is connected. Therefore $D(a)$ is connected.
\endproof

\section{Binary-inputs MAC} \label{S:BinaryMAC}

In this section, we investigate an $N$-user binary-inputs MAC $(P,Y)$ of
the $N$-user $(n_1,\cdots,n_N;m)$-MAC $(P,X)$ where each $Y_k$ of $Y$ is
formed by a line segment. For any given $\fat{\rho}_k',\fat{\rho}_k''\in
X_k$, $k=1,\cdots,N$, define a line segment $Y_k$ by
\[
Y_k=\{\theta_k\fat{\rho}_k'+(1-\theta_k)\fat{\rho}_k''|0\leq\theta_k\leq
1\},\ \  k=1,\cdots,N
\]
and denote $Y$ $=$ $Y_1\times\cdots\times Y_N$. Reasonably we set
$\fat\theta \equiv ({\theta}_1,\cdots,{\theta}_N)$ and write ${\theta}_k
\in Y_k$, $\fat\theta \in Y$. Thus we can build up an $N$-user
binary-inputs $(2,\cdots,2;m)$-MAC $(P,Y)$ whose channel matrix is $P$ and
domain is a subset $Y$ of $X$. Obviously it is an elementary MAC since
$m\geq 2$.

The mutual information of the $N$-user $(2,\cdots,2;m)$-MAC $(P,Y)$ is
given by
\begin{eqnarray}
\lefteqn{{\cal I}(\theta_1,\cdots,\theta_N;\fat{\rho}',\fat{\rho}'')\equiv
I((\theta_1\fat{\rho}_1'+(1-\theta_1)\fat{\rho}_1'')\times} \nonumber \\
&& \cdots
\times(\theta_N\fat{\rho}_N'+(1-\theta_N)\fat{\rho}_N''))\hskip5em
\label{EQ:NON-SUB}
\end{eqnarray}
where $0\leq\theta_k\leq 1$, $k=1,\cdots,N$, and
$\fat{\rho}'=\fat{\rho}_1'\times\cdots\times\fat{\rho}_N'$,
$\fat{\rho}''=\fat{\rho}_1''\times\cdots\times\fat{\rho}_N''$. It depends
on the choice of $\fat{\rho}',\fat{\rho}''$. The Kuhn-Tucker conditions
for (\ref{EQ:NON-SUB}) are given by
\begin{eqnarray}
{\cal I}_k(\theta_1,\cdots,\theta_N;\fat{\rho}',\fat{\rho}'')&=&
0,\ \ \theta_k>0 \nonumber \\
&\leq& 0,\ \ \theta_k=0 \nonumber \\
&& k=1,\cdots,N \label{EQ:BIN-KT}
\end{eqnarray}
where ${\cal I}_k(\cdots;\fat{\rho}',\fat{\rho}'')=$
${\partial}{\cal{I}}(\cdots;\fat{\rho}',\fat{\rho}'') /
{\partial\theta_k}$. For simplicity we omit $\fat{\rho}',\fat{\rho}''$
from the expression and denote ${\cal
I}(\fat\theta;\fat{\rho}',\fat{\rho}'') \equiv {\cal I}(\fat\theta)$, in
the subsequent discussions.

We prove the lemma to be used for the proof of Theorem~\ref{T:E-MAC} as
follows:
\begin{lemma} \label{L:BIN-MAC}
The Kuhn-Tucker conditions for the $N$-user binary $(2,\cdots,2;m)$-MAC
$(P,Y)$ as defined above are necessary and sufficient for optimality.
\endtheorem
\end{lemma}

\proof It is sufficient to prove the sufficiency. Assume that there exist
two solutions $\bar{\fat{\theta}}=(\bar{\theta}_1,\cdots,\bar{\theta}_N)$
and $\hat{\fat{\theta}}=(\hat{\theta}_1,\cdots,\hat{\theta}_N)$ of the
Kuhn-Tucker conditions (\ref{EQ:BIN-KT}) such that ${\cal
I}(\bar{\fat{\theta}})\neq {\cal I}(\hat{\fat{\theta}})$. Without loss of
generality, assume that ${\cal I}(\bar{\fat{\theta}})>{\cal
I}(\hat{\fat{\theta}})$.

Since the $N$-user binary $(2,\cdots,2;m)$-MAC $(P,Y)$ is elementary, by
Proposition~\ref{P:MAXIMAL} the solution $\hat{\fat{\theta}}$ is local
maximum in $Y$ and there exists a neighborhood $U_{\hat{\sfat{\theta}}}$
of $\hat{\fat{\theta}}$ such that ${\cal I}(\hat{\fat{\theta}})\geq {\cal
I}(\fat{\theta}\in U_{\hat{\sfat{\theta}}})$. Also by
Proposition~\ref{P:CONNECTED} the set $D({\cal
I}(\hat{\fat{\theta}}))\equiv$ $\{\fat{\theta}|{\cal
  I}(\fat{\theta}){\geq}
{\cal I}(\fat{\hat{\theta}}),\fat{\theta}\in Y\}$ is connected and
includes both $\bar{\fat{\theta}}$ and $\hat{\fat{\theta}}$. Then for any
$\fat{\theta}\in D({\cal I}(\hat{\fat{\theta}}))\cap
 U_{\hat{\sfat{\theta}}}$, it is easy to see
${\cal I}(\fat{\theta})= {\cal I}(\hat{\fat{\theta}})$.

Let $\fat{\theta}^*$ and $\fat{\theta}^\dagger$ be any points in $D({\cal
I}(\hat{\fat{\theta}}))\cap U_{\hat{\sfat{\theta}}}$, and set ${\cal
I}(\theta_1^*,\cdots,(\alpha\theta_k^*+(1-\alpha)\theta_k^\dagger),
\cdots,\theta_N^*)$ as a function of the variable $\alpha$. Since ${\cal
I}(\fat\theta)$ is concave for each variable $\theta_k$ and ${\cal
I}(\fat{\theta}^*)= {\cal I}(\fat{\theta}^\dagger)=I(\hat{\fat{\theta}})$,
we have ${\cal
I}(\theta_1^*,\cdots,(\alpha\theta_k^*+(1-\alpha)\theta_k^\dagger),
\cdots,\theta_N^*)$ $=$ ${\cal I}(\hat{\fat{\theta}})$ for
$0\leq\alpha\leq 1$. Therefore, it holds
\begin{eqnarray*}
\lefteqn{\frac{d{\cal I}(\theta_1^*,\cdots,(\alpha\theta_k^*
+(1-\alpha)\theta_k^\dagger),
\cdots,\theta_N^*)}{d\alpha}=} \\
&&(\theta_k^*-\theta_k^\dagger) {\cal
I}_k(\theta_1^*,\cdots,(\alpha\theta_k^*+(1-\alpha)\theta_k^\dagger),
\cdots,\theta_N^*) \\
&&=0.
\end{eqnarray*}
This implies that any $\fat{\theta}\in D({\cal I}(\hat{\fat{\theta}}))
\cap U_{\hat{\sfat{\theta}}}$ satisfies the Kuhn-Tucker conditions
(\ref{EQ:BIN-KT}): ${\cal
I}_k(\theta_1,\cdots,\theta_k,\cdots,\theta_N)=0$, $k=1,\cdots,N$, even if
$\hat{\fat{\theta}}$ is located on the boundary of $Y$.

\begin{figure}[t]
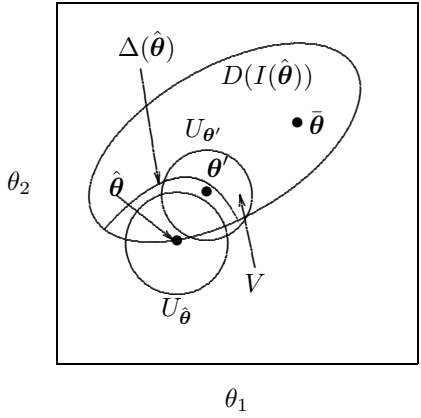

\centerline{%
 \beginpicture
\setcoordinatesystem units <.8mm,.8mm> \setplotarea x from 0 to 60, y from
0 to 60 \axis left label {$\theta_2$} ticks length <0pt> / \axis bottom
label {$\theta_1$} ticks length <0pt> / \axis top ticks length <0pt> /
\axis right ticks length <0pt> / \put {$\bullet$} at 20 20.4 \put
{$\hat{\fat{\theta}}$} at 10 30 \arrow <4pt> [.2,.5] from 10 28 to 19.4
20.6 \put {$\bullet$} at 40 40 \put {$\bar{\fat{\theta}}$} [l] at 42 40
\put {$D(I(\hat{\fat{\theta}}))$} at 35 48 \put
{$U_{\hat{\sfat{\theta}}}$} [t] at 20 11 \ellipticalarc axes ratio 1:1 360
degrees from 14 14 center at 20 20 \put {$\bullet$} at 25 28.5 \put
{$\fat{\theta}'$} [lb] at 25 31 \put {$U_{\sfat{\theta}'}$} [b] at 24.5 37
\ellipticalarc axes ratio 1:1 360 degrees from 25 35.5 center at 25 28
\setquadratic \plot 8 22.4  21 31   30.2 23.7 / \put
{$\Delta(\hat{\fat{\theta}})$} [b] at 15 50 \arrow <4pt> [.2,.5] from 14
49 to 17 30
\put {$V$} [t] at 33 15 \arrow <4pt> [.2,.5] from 33 16 to 30.5 28
\rotatebox{30} {\ellipticalarc axes ratio 2:1 360 degrees from 26 15.5
center at 25 28 }
\endpicture
} \caption{Two solutions of the Kuhn-Tucker conditions for two-user case.}
 \label{F:2SOL-KTC}
\end{figure}

Let $\Delta(\hat{\fat{\theta}})$ be a set
\begin{eqnarray*}
\lefteqn {\Delta(\hat{\fat{\theta}})\equiv \{{\fat{\theta}|{\cal
I}_k(\fat\theta)=0,} \ \ k=1,\cdots,N,} \nonumber \\
&& {\cal I}(\fat{\theta}) = {\cal I}(\hat{\fat{\theta}}), \ \
\fat{\theta}\in{}D(I(\hat{\fat{\theta}}))\}.
\end{eqnarray*}
Clearly, this includes $D({\cal I}(\hat{\fat{\theta}}))\cap
U_{\hat{\sfat{\theta}}}$ and it holds ${\cal I}(\fat{\theta})= {\cal
I}(\hat{\fat{\theta}})$ for $\fat{\theta}\in\Delta(\hat{\fat{\theta}})$
(see Fig~\ref{F:2SOL-KTC}). Note that each point in
$\Delta(\hat{\fat{\theta}})$ is local maximum. Then for any
$\fat{\theta}'\in\Delta(\hat{\fat{\theta}})$, there exists a neighborhood
$U_{\sfat{\theta}'}$, such that ${\cal I}(\fat{\theta})\leq {\cal
I}(\fat{\theta}') (={\cal I}(\hat{\fat{\theta}}))$ for any
$\fat{\theta}\in U_{\sfat{\theta}'}$.

Here we define a subset of $U_{\sfat{\theta}'}$ as
$V\equiv\{\fat{\theta}|\fat{\theta}\in U_{\sfat{\theta}'},
\fat{\theta}\not\in\Delta(\hat{\fat{\theta}})\}\cap D({\cal
I}(\hat{\fat{\theta}}))$. Assume that $V$ is non-empty. Then it holds
${\cal I}(\fat{\theta})< {\cal I}(\hat{\fat{\theta}})$ for any
$\fat{\theta}\in V$, since $\fat{\theta}\in\{\fat{\theta}|\fat{\theta}\in
U_{\sfat{\theta}'}, \fat{\theta}\not\in\Delta(\hat{\fat{\theta}})\}$. On
the other hand, it holds that ${\cal I}(\fat{\theta})\geq {\cal
I}(\hat{\fat{\theta}})$ for any $\fat{\theta}\in V$ since $\fat{\theta}\in
D({\cal I}(\hat{\fat{\theta}}))$ by the definition of $V$. This is
inconsistent with the assumption that $V$ is non-empty. Thus $V$ is empty
and $\Delta(\hat{\fat{\theta}})=D({\cal I}(\hat{\fat{\theta}}))$.

Since both $\hat{\fat{\theta}}$ and $\bar{\fat{\theta}}$ belong to
$D({\cal I}(\hat{\fat{\theta}}))$, it holds ${\cal I}(\hat{\fat{\theta}})=
{\cal I}(\bar{\fat{\theta}})$.  Therefore the assumption ${\cal
I}(\bar{\fat{\theta}})> {\cal I}(\hat{\fat{\theta}})$ is invalid. This
means that any solution $\fat\theta$ of the Kuhn-Tucker conditions
(\ref{EQ:BIN-KT}) for $(P,Y)$ gives the same value for ${\cal
I}(\fat\theta)$and then it is optimal. Thus the sufficiency is
proved.\endproof

Note that the Lemma~\ref{L:BIN-MAC} holds for any domain $Y$ of $X$ formed
by $\fat{\rho}'$ and $\fat{\rho}''$.

\section{Proof of Theorem~\ref{T:E-MAC}} \label{S:PROOF}
In this section, we prove Theorem~\ref{T:E-MAC} by using
Lemma~\ref{L:BIN-MAC}. We state again Theorem~\ref{T:E-MAC}:

{\it Theorem~\ref{T:E-MAC}:} The Kuhn-Tucker conditions for the channel
capacity $C$ of an $N$-user elementary $(n_1,\cdots,n_N;m)$-MAC $(P,X)$,
where $n_k\leq m$ for all $k=1,\cdots,N$, are necessary and sufficient.
\endtheorem

\proof It is sufficient to prove the sufficiency. Let
$\bar{\fat{p}}=\bar{\fat{p}}_1\times\cdots\times\bar{\fat{p}}_N$ a
solution of the Kuhn-Tucker conditions (\ref{EQ:MAC-KT}) for the $N$-user
elementary $(n_1,\cdots,n_N;m)$-MAC $(P,X)$, where $n_k\leq m$,
$k=1,\cdots,N$. We prove that $\bar{\fat{p}}$ is uniquely determined in
the sense that any solution $\fat{p}$ of the Kuhn-Tucker conditions
(\ref{EQ:MAC-KT}) gives the same for $I(\fat{p})$.

For an arbitrary $\fat{p}_k'\in X_k$, $k=1,\cdots,N$, there exist
$\fat{p}_k''\in X_k$ and $\bar{\theta}_k$ such that
\begin{equation}
\bar{\fat{p}}_k=\bar{\theta}_k\fat{p}_k'+(1-\bar{\theta}_k)\fat{p}_k'', \
\ 0\leq\bar{\theta}_k\leq 1 \label{EQ:theta_bar}
\end{equation}
since $X_k$ is simplex. Then $\bar{\fat{p}}$ is represented by
\begin{eqnarray}
\lefteqn{\bar{\fat{p}}=
(\bar{\theta}_1\fat{p}_1'+(1-\bar{\theta}_1)\fat{p}_1'')\times}\nonumber \\
&&\cdots \times(\bar{\theta}_N\fat{p}_N'+(1-\bar{\theta}_N)\fat{p}_N'').
\label{EQ:barfatp}
\end{eqnarray}

Here we define a function of variables ($\theta_1, \cdots,\theta_N$)
$\equiv \fat\theta$ ($0\leq\theta_k\leq 1$) by
\begin{eqnarray}
\lefteqn{{\cal I}(\fat\theta;\fat{p}',\fat{p}'')\equiv
I((\theta_1\fat{p}_1'+(1-\theta_1)\fat{p}_1'')\times} \nonumber \\
&& \cdots \times(\theta_N\fat{p}_N'+(1-\theta_N)\fat{p}_N'')) \hskip5em
\label{EQ:cal_I}
\end{eqnarray}
where $\fat{p}'=\fat{p}_1'\times\cdots\times\fat{p}_N'$ and $\fat{p}''
=\fat{p}_1''\times\cdots\times\fat{p}_N''$. The function (\ref{EQ:cal_I})
can be regarded as the mutual information of an $N$-user
$(2,\cdots,2;m)$-MAC $(P,Y)$ with the domain $Y\equiv
Y_1\times\cdots\times Y_N$, where $Y_k\equiv \{\theta_k
\fat{p}_k'+(1-\theta_k)\fat{p}_k''|0\leq\theta_k\leq{1}\}$,
$k=1,\cdots,N$. The $N$-user $(2,\cdots,2;m)$-MAC $(P,Y)$ is denoted by
$(P,Y)_{(\sfat{p}',\sfat{p}'')}$, since it depends on $\fat{p}',
 \fat{p}''$.

Since $\bar{\fat{p}}$ is a solution of the Kuhn-Tucker conditions
(\ref{EQ:MAC-KT}) for $(P,X)$, then $\fat{\bar\theta}$ is a solution of
the Kuhn-Tucker conditions for the mutual information (\ref{EQ:cal_I}) of
$(P,Y)_{(\sfat{p}',\sfat{p}'')}$:
\begin{eqnarray}
{\cal I}_k(\fat\theta;\fat{p}',\fat{p}'')&=&
{\cal C},\ \ \theta_k>0 \nonumber \\
&\leq& {\cal C},\ \ \theta_k=0 \nonumber
\end{eqnarray}
\begin{equation}
k=1,\cdots,N, \ \ {\cal C}={\cal I}(\fat\theta;\fat{p}',\fat{p}'')
\label{EQ:KT(P,Y)}
\end{equation}
where ${\cal I}_k(\fat\theta;\fat{p}',\fat{p}'') =
{\partial}{\cal{I}}(\fat\theta;\fat{p}',\fat{p}'')/{\partial\theta_k}$.
Therefore, it follows from Lemma~\ref{L:BIN-MAC} that $\fat{\bar\theta}$
is optimal for $(P,Y)_{(\sfat{p}',\sfat{p}'')}$, which means
\begin{equation}
{\cal{I}}(\fat{\bar\theta};\fat{p}',\fat{p}'')\geq
{\cal{I}}(\fat{\theta};\fat{p}',\fat{p}'') \label{EQ:A1}
\end{equation}
for any $\fat{\theta}\in Y$.

Since $\fat{\bar\theta}$ is given by (\ref{EQ:barfatp}), it holds
\begin{equation}
{\cal{I}}(\fat{\bar\theta};\fat{p}',\fat{p}'')
 =I(\bar{\fat{p}}) \label{EQ:A2}
\end{equation}
for any $\fat{p}'\in X$, where $\fat{p}''$ satisfies (\ref{EQ:theta_bar}).
Thus since (\ref{EQ:A1}) and (\ref{EQ:A2}) are valid for any $\fat{p}'\in
X$, it holds
\[
I(\bar{\fat{p}})\geq I(\fat{p})
\]
on the whole domain $X$. This implies that $\bar{\fat{p}}$ is optimal.

Thus we proved the theorem.
\endproof

\section{Conclusions}\label{S:CONCLUTIONS}

After Shannon~\cite{Two-Way} multiuser channel has long been studied
in various fields. However not much works have been made for the
fundamental property of the channel capacity of an $N$-user
$(n_1,\cdots,n_N)$-MAC $(P,X)$ in general except for some specific
cases.

We have shown that there exists a non-trivial MAC where the
Kuhn-Tucker conditions are necessary and sufficient for the channel
capacity. We called it as an \emph{elementary} MAC that was defined
by the MAC whose sizes of input alphabets must be not greater than
the size of output alphabet. Obviously the $N$-user binary inputs
$(2,\cdots,2;m)$-MAC $(P,X)$ is a typical example of the elementary
MAC. Also the DMC is a trivial elementary MAC.

We believe that there is considerable merit in a concept of elementary MAC
for which the channel capacity is evaluated precisely by the necessary and
sufficient condition as in the case of DMC. In fact, we have proved as
Theorem~\ref{T:C(F)} that the channel capacity of any MAC is achieved by
the channel capacity of \emph{an} elementary MAC contained in the original
MAC. Thus an MAC in general can be regarded as simply an aggregate of
elementary MAC's. This statement is a basic idea behind our formulation of
this paper.

The most of this paper was devoted to the proof of
Theorem~\ref{T:E-MAC} such that the Kuhn-Tucker conditions are
sufficient (the necessity is self-evident) for the channel capacity
of the elementary MAC. We have shown as Proposition~\ref{P:KT=BEQ}
that a solution of the Kuhn-Tucker conditions if it satisfies the
equality portion of the conditions satisfies the boundary equations
which define the boundary of the capacity region. Then we could
prove the property of local maximum as Proposition~\ref{P:MAXIMAL}
followed by the property of connectedness as
Proposition~\ref{P:CONNECTED}. By using these two distinctive
features we could prove that any solution of the Kuhn-Tucker
conditions of the elementary MAC was uniquely determined, that is,
each solution takes the same value for the the mutual information
and therefore it achieves the channel capacity.

In this respect, we remark that the non-elementary MAC has a
degenerate property as explained in Section~\ref{S:E-MAC}. If it
exists, then it is difficult to identify which IPD vectors are
exactly contributed to the mutual information of the MAC. However we
overcome these difficulties by introducing the concept of elementary
MAC where there exists no such degenerate property. Since the
well-known DMC is elementary, then the elementary MAC is identified
as an extension of the DMC.

Incidentally, our notation introduced in this paper seems rather
non-standard including expressions of IPD vector $\fat p$, Kronecker
products $\fat{p} = \fat{p}_1\times\cdots\times \fat{p}_N$, the channel
matrix $P$ regarded as a non-linear mapping, domain $X$, face $F$, and so
force. However we emphasize that the notation appears effective to resolve
the cumbersome procedures relating to the extremum evaluation of the
multi-variable mutual information with constraints for the MAC.

Before closing we remark that the very essence of information theory
consists in two major subjects such as source coding and channel
coding as we know. This paper seems to be quite effective in working
out the subject of channel coding since we provide for a formalism
to determine the channel capacity of the MAC. We are confident that
two distinctive features of local maximum
(Proposition~\ref{P:MAXIMAL}) and connectedness
(Proposition~\ref{P:CONNECTED}) represent an intrinsic structure of
the MAC. However we are not content ourselves with this stage. We
are expecting that our results will be a mathematical base for
various subjects of the MAC including the numerical and/or exact
evaluation of the capacity region, the analysis of the MAC with
feedback as well as the structured approach to the multiuser coding,
and so force.

\end{document}